\documentclass[amssymb,amsmath,aps,prb,showpacs,twocolumn]{revtex4}
\usepackage{times}
\usepackage{graphicx}
\usepackage{bm}
\usepackage{latexsym}
\begin{document}
\title{Spin-Polarized Electron Transport at Ferromagnet/Semiconductor Schottky Contacts}
\author{J. D. Albrecht}
\affiliation{Air Force Research Laboratory, Wright-Patterson AFB,
Ohio 45433}
\author{D. L. Smith}
\affiliation{Los Alamos National Laboratory, Los Alamos, New
Mexico 87545}

\begin{abstract}
We theoretically investigate electron spin injection and
spin-polarization sensitive current detection at Schottky contacts
between a ferromagnetic metal and an n-type or p-type
semiconductor.  We use spin-dependent continuity equations and
transport equations at the drift-diffusion level of approximation.
Spin-polarized electron current and density in the semiconductor
are described for four scenarios corresponding to the injection or
the collection of spin polarized electrons at Schottky contacts to
n-type or p-type semiconductors.  The transport properties of the
interface are described by a spin-dependent interface resistance,
resulting from an interfacial tunneling region.  The
spin-dependent interface resistance is crucial for achieving spin
injection or spin polarization sensitivity in these
configurations.  We find that the depletion region resulting from
Schottky barrier formation at a metal/semiconductor interface is
detrimental to both spin injection and spin detection.  However,
the depletion region can be tailored using a doping density
profile to minimize these deleterious effects.  For example, a
heavily doped region near the interface, such as a delta-doped
layer, can be used to form a sharp potential profile through which
electrons tunnel to reduce the effective Schottky energy barrier
that determines the magnitude of the depletion region.  The model
results indicate that efficient spin-injection and
spin-polarization detection can be achieved in properly designed
structures and can serve as a guide for the structure design.
\end{abstract}

 \maketitle

\section{introduction}
Semiconductor physics is in the midst of a wide ranging
exploration of physical phenomena and device concepts that are
connected with the electron spin degree of freedom.  Much work has
focused on optical generation and detection of spin populations.
However, most spin based device concepts require an electrical
means of injecting, manipulating, and detecting spin-polarized
electron currents.  Thus it is important to understand the
fundamental physics of electron spin transport in the main
structural components that make up semiconductor devices.  The
Schottky contact is an essential semiconductor device component.
Schottky contacts form at most metal/semiconductor interfaces.
Electrical spin injection and detection schemes often involve
ferromagnetic metal/semiconductor interfaces with Schottky
contacts so it is important to understand spin dependent electron
transport at these structures.

Presently, theories for the injection or detection of
spin-polarized electron currents at a metallic
ferromagnet/nonmagnetic semiconductor interface have been
formulated in the spirit of transport at a ferromagnetic
metal/normal metal interface.  The description of spin transport
is incorporated using variations on a spin diffusion
equation.\cite{wyder}  In these
approaches\cite{schmidt,rashba,smithsilver,pikus,yu,rashba2} the
semiconductor is described as a poorly conducting metal, in the
sense that the the carrier density and thus the conductivity of
the semiconductor is taken to be spatially uniform.  Important
insights gained through these models include: ({\it i}) the large
conductivity mismatch between a highly conductive metal and a
comparatively weakly conductive semiconductor is a major obstacle
to spin injection; and ({\it ii}) a spin selective interface
resistance can be of great benefit to efficient spin injection.  A
major drawback to a spin device physics model based on such
uniform conductivity treatments is that they do not describe the
underlying electronic properties, the currents and potentials of
real semiconductor structures.  An obvious example of a problem
with uniform conductivity models for metal/semiconductor Schottky
contacts is that they yield the symmetric, linear current-voltage
characteristics of resistors rather than the rectifying
characteristics of diodes.  An initial study of spin injection
including the effects of band-bending in a depletion region at an
n-type Schottky contact showed that the depletion region can have
an important effect on spin transport and that a device-physics
approach to the theory of spin-contacts is necessary.\cite{brief}

Experimentally, spin dependent transport has been investigated at
interfaces consisting of a ferromagnetic
metal\cite{ploogmetal,crowell,jonkermetal} or a heavily doped
spin-polarized semiconductor\cite{jonkerhet,molenhet} contact and
a non-magnetic semiconductor.  Both spin injection, in which the
electron flux flows from the spin polarized contact into the
nonmagnetic semiconductor, and spin detection, in which the
electron flux flows from the nonmagnetic semiconductor into the
spin polarized contact, have been considered.  In the spin
injection measurements, detection of spin-polarized injection is
often made using a spin-LED (light-emitting diode) configuration.
In these experiments, electrons are injected into an n-type
semiconductor from a spin polarized contact and are subsequently
transported to a detection region, typically a quantum well, where
they recombine with unpolarized holes transported from an adjacent
p-type doped region.  Because of the optical selection rules in
III-V semiconductors the relative intensity of right- and left-
circularly polarized luminescence gives a measure of the
spin-polarization of the electron density in the recombination
region.  In spin detection measurements, spin polarized electrons
are often optically generated in III-V semiconductors, and a spin
dependent voltage signal is sought as the electron flux is
transported into a spin polarized contact.

In this paper, we theoretically investigate spin-polarized
electron current at ferromagnetic metal/semiconductor Schottky
contacts.  We systematically treat the semiconductor device
operation and the spin physics at the same level of approximation.
We consider both n-type and p-type Schottky contacts with current
flow corresponding to either forward and reverse bias.  We first
treat the overall electrostatics of the system and subsequently
solve charge and spin continuity equations.  We use a
drift-diffusion transport model to describe the charge and spin
currents.  The drift-diffusion transport model is a strong
scattering approximation appropriate for relatively high
temperatures, such as room temperature.  It is the approach used
to describe most semiconductor device operation.  Here we extend
this this approach to describe spin dependent transport at
Schottky contacts.  We find that the depletion region associated
with a Schottky energy barrier can have a very strong effect on
spin-polarized electron transport at ferromagnetic
metal/semiconductor contacts.  A large Schottky barrier is
detrimental to spin injection and can also hinder spin detection.
The model suggest structure design strategies for reducing the
detrimental effects of the Schottky energy barrier.

The paper is organized in the following way: in Sec. II we
describe the model, in Sec. III we present our numerical results
and in Sec. IV we summarize our conclusions.  Calculational
details are included in the appendices.

\section{description of model}
When a metal/semiconductor interface is formed the Fermi energy is
usually pinned within the energy gap of the semiconductor.  The
position of the the semiconductor valence and conduction bands,
relative to Fermi energy, at the interface does not depend
strongly on the bulk doping of the semiconductor or on which metal
is used to make the contact.  For a given semiconductor, this
energy matching position at the interface is largely fixed.
Generally the position of the semiconductor valence and conduction
bands relative to the Fermi energy at the interface, which depends
on interfacial charge distribution, does not coincide with the
corresponding energy position of the bands in the bulk of the
semiconductor, which depends on the bulk doping level.  There is a
band bending region near the interface which at zero applied bias:
is depleted of carriers, is charged because of the background
doping, and has a large spatially varying electric field.  The
Schottky energy barrier between the pinned Fermi level and the
semiconductor conduction band at the interface results in the
charged depletion region and has important consequences on charge
current flow at metal/semiconductor interfaces.  For example, it
leads to diode type current-voltage characteristics.  Thus, it is
not particularly surprising that this energy barrier and depletion
region also have important consequences on spin current flow at
these interfaces.

The design of the interface is central to spin injection and
detection structures.  In particular, spin dependent interface
resistance resulting from spin dependent tunnel barriers have been
argued to be essential for effective spin injection or detection
at metal semiconductor interfaces.
\cite{schmidt,rashba,smithsilver}  Possible spin-selective
interface resistance layers, formed from thin magnetic insulators,
have been experimentally investigated by Motsnyi, et
al.\cite{motsnyi}  In other work, Hanbicki, et al., have
investigated Schottky barriers with heavy doping near the
interface to study structures in which current is dominated by
tunneling for spin injection.\cite{jonkerdep}  These results are
promising for the realization of future electron spin based device
designs.  Interfacial spin-flip scattering, which would be
detrimental to spin injection or detection structures, is
possible.\cite{interface}  Structures should be designed to
minimize this process.

\begin{figure*}
\rotatebox{0}{ \includegraphics{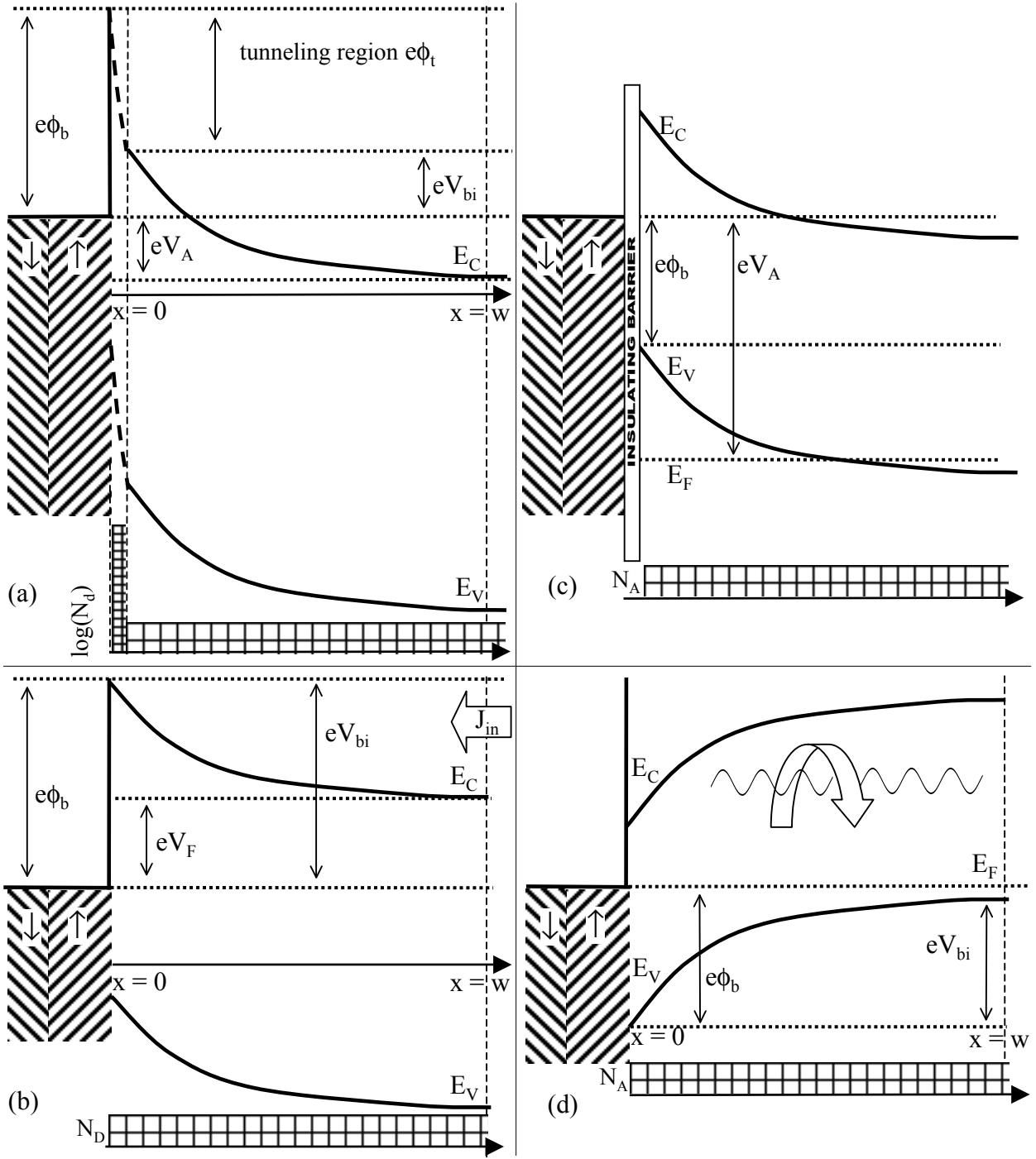}} \caption{\label{fig1}
Energy diagram of a Schottky contact for four cases: (a) electron
spin injection into an n-type semiconductor, (b) spin current
detection from an n-type semiconductor, (c) spin injection into an
accumulated p-type semiconductor, and (d) spin current detection
from an optically polarized p-type semiconductor.}
\end{figure*}

We consider four scenarios corresponding to the injection or
collection of spin polarized electron current at Schottky contacts
to n-type or p-type semiconductors.  The four cases are
schematically shown in Fig.\ \ref{fig1}.  Panel (a) of Fig.\
\ref{fig1} illustrates the case of spin injection into a n-type
semiconductor.
The diode formed by the Schottky contact is in reverse bias and
the electron flux is from the ferromagnetic metal on the left into
the semiconductor on the right.  A heavily doped region near the
interface, as illustrated by the doping profile in the lower part
of panel (a), can be designed to form a sharp potential profile
through which electrons tunnel.  The heavily doped region reduces
the effective Schottky energy barrier that determines the
properties of the depletion region.\cite{sze}  The total barrier
e$\phi _{b}$ is divided into two parts, a tunneling region with
barrier height e$\phi _{t}$ and an effective Schottky barrier
height $eV_{bi}$.  The potential drop in the depletion region
consists of the effective Schottky barrier height plus the applied
reverse bias $eV_{R}$.  Two parameters of the tunneling region,
its tunneling resistance and the magnitude of the reduction of the
effective Schottky barrier, can be separately controlled by the
parameters of the doping profile, for example the height and width
of the heavily doped region.

Panel (b) of Fig.\ \ref{fig1} illustrates the case of
spin-polarization sensitive current detection at a Schottky
contact between a ferromagnetic metal and an n-type semiconductor.
A spin-polarized electron flux is incident from the semiconductor
and the Schottky diode is in forward bias.  In a typical
experimental situation, the structure is held under constant
current bias and a change in voltage signal is sought when the the
polarity of the spin-polarized incident current is reversed. There
may be a heavily doped region near the interface as in panel (a).

Panel (c) of Fig.\ \ref{fig1} illustrates the case of electron
spin injection into a p-type semiconductor.  The p-type Schottky
diode is in strong forward bias.  There is an insulating tunneling
barrier at the interface that limits the hole current, which
nonetheless can be considerable.  There is a hole accumulation
region in the semiconductor near the interface.  The minority
carrier electron flux is from the ferromagnetic metal into the
semiconductor.  This structure can be interesting for
characterizing the spin dependent transport properties of the
tunneling barrier.

Panel (d) of Fig.\ \ref{fig1} illustrates the case of
spin-polarization sensitive current detection at a Schottky
contact between a ferromagnetic metal and a p-type semiconductor.
The p-type Schottky diode is under zero or small (either forward
or reverse) bias.  Spin polarized electrons are optically
generated by absorption of circularly polarized light.  In a
typical experimental situation, the structure is held under
constant current bias and a fixed incident optical intensity and a
change in voltage signal is sought when the the polarity of the
circularly polarized incident light is reversed.  This is the
essentially the same structure as in panel (c), except under
different bias and optical excitation conditions.  These two
experimental configurations can be used together to characterize
the spin transport properties of a tunneling barrier at the
ferromagnet/semiconductor interface.

We describe the ferromagnetic metal and the interface using a spin
dependent drift-diffusion equation, a spin diffusion equation, and
spin dependent interface conductances as in Ref.
\onlinecite{smithsilver}.  The drift-diffusion equation describing
current flow in the ferromagnetic metal is
\begin{equation}\label{eqn1}
j_{\eta }=\sigma _{\eta }\frac{\partial \left( \mu _{\eta }/e\right) }{%
\partial x}
\end{equation}
Here $j_{\eta }$ is the current density due to electrons of spin
type $\eta$$($$=$$\uparrow$,$\downarrow$$)$, $\sigma _{\eta }$ is
the conductivity for electrons of that spin type, $\mu _{\eta }$
is the corresponding electro-chemical potential, $e$ is the
magnitude of the electron charge and $x$ is position.  Eq.\
(\ref{eqn1}) assumes rapid wave vector randomizing scattering
events, so that electrons of the same spin stay in local
quasi-thermal equilibrium with each other.  However, spin-flip
scattering can be comparatively slow so that electrons of
different spin may be driven out of local quasi-thermal
equilibrium by, for example, an applied current density.  When
electrons with different spins are driven out of local
quasi-thermal equilibrium, so that $\mu _{\uparrow }$ is not equal
to $\mu _{\downarrow}$ at some point in space, spin relaxation
away from that spatial point is described by a diffusion equation
\begin{equation}\label{eqn2}
\frac{{\partial ^2  {\mu_-} }}{{\partial x^2 }} = \frac{{ {\mu_-}
}}{{\Lambda_c ^2 }}.
\end{equation}
Here $\Lambda_c$ is the spin diffusion length in the metallic
contact and we use the notation
$\mu_\pm$$=$$\mu_\uparrow$$\pm$$\mu_\downarrow$.  At the
contact/semiconductor interface, electrons of different spin can
be driven out of quasi-thermal equilibrium by current flow.  Far
from the interface, as $x$$\rightarrow$$\pm \infty$, the
electrochemical potential difference vanishes
$\mu_-$$\rightarrow$$0$.  The total steady state current density
is a constant function of position.  We assume no strong spin flip
scattering at the interface so that the individual current
components for the two spin types are continuous at the interface.
Current flow at the interface is described using an interface
resistance
\begin{equation}\label{eqn3}
j_\eta ^o  =  \frac{\Delta \mu_\eta}{e R_\eta}
\end{equation}
where $j_\eta ^o$ is the current density at the interface,
$R_\eta$ is the interface resistance, and $\Delta \mu_\eta$ is an
interfacial discontinuity in electro-chemical potential for
electrons of spin type $\eta$.  If the interface resistance is
zero, the electro-chemical potentials are continuous at the
interface whereas for nonzero values of $R_\eta$ a discontinuity
in $\mu_\eta$ can develop at the interface.  For notational ease,
we set the variables $j_\pm$$=$$j_\uparrow$$\pm$$j_\downarrow$. In
the contact where the hole current is zero, the total current
$j$$=$$j_+$.  We take the electron density as a function of
position to be fixed, independent of the current density $j$, in
the contact.  The total conductivity of the contact is then
independent of position and current density.  It is convenient to
define a contact polarization variable $\alpha_c$ by
$\sigma_\uparrow$$=$$\alpha_c \sigma_c$ or
$\sigma_\downarrow$$=$$($$1$$-$$\alpha_c$$)$$\sigma_c$ where
$\sigma_c$ is the total contact conductivity.

We take the contact on the left ($x$$<$$0$) and the semiconductor
on the right ($x$$>$$0$) of the interface located at $x=0$, as in
Fig.\ \ref{fig1}, so that the current density is negative for
electron injection into the semiconductor.  Solving Eq.\
(\ref{eqn2}) with the stated boundary conditions gives
\begin{equation}\label{eqn4}
\mu_- = \mu_-^{0^-}  e^ { x/ \Lambda_c}  \qquad\hbox{for $x < 0$}.
\end{equation}
Quantities evaluated at the interface approached from the contact
and semiconductor are indicated by the superscripts $0^-$ and
$0^+$, respectively.  From Eqs.\ (\ref{eqn1}) and (\ref{eqn4}) we
find
\begin{eqnarray}\label{eqn5}
\mu_-^{0^-}&=&e\Lambda_c \left( \frac {j_ \uparrow ^{0^-}}{\sigma
_ \uparrow} - \frac {j_ \downarrow ^{0^-}}{\sigma _ \downarrow}
\right) \nonumber \\
&=&\frac {e\Lambda_c}{2\sigma_c \alpha_c \left ( 1-\alpha_c \right
) } \left( j_+^{0^-}+j_-^{0^-} -2\alpha_c j_+^{0^-}\right),
\end{eqnarray}
and
\begin{equation}\label{eqn6}
\frac {\partial \mu_+}{\partial x} = \frac {2ej} {\sigma_c} +
\frac {\left( 1 - 2 \alpha_c \right)} {\Lambda_c} \mu_-^{0^-} e^
{x/\Lambda_c}.
\end{equation}

The total current in the semiconductor is
$j$$=$$j_\uparrow$$+$$j_\downarrow$$+$$j_p$$=$$j_+$$+$$j_p$ where
$j_p$ is the hole current density.  For n-type Schottky barriers
$j_p$$=$$0$, but not necessarily for the p-type structures.  It is
convenient to define $\beta$ as the fraction of the electron
current carried by spin up electrons
$\beta$$=$$j_\uparrow$$/$$j_+$.  We assume that there is not
strong spin-flip scattering at the interface so that
$j_-^{0^-}$$=$$j_-^{0^+}$.  The total current $j$ is continuous at
the interface.  The interface resistance conditions, Eq.\
(\ref{eqn3}) then lead to the interface matching conditions,
\begin{multline}\label{eqn7}
\mu_-^{0^-} = \frac {e j_+^{0^+} \Lambda_c}{\sigma_c
\alpha_c
(1-\alpha_c)} \\
\times \left [ \beta^{0^+} - \alpha_c - \left ( \alpha_c -
{\textstyle{1 \over 2}} \right ) \frac {\left ( j-j_+^{0^+}
\right)} {j_+^{0^+}} \right ]
\end{multline}
\begin{equation}\label{eqn8}
\mu_-^{0^+} - \mu_-^{0^-} = e j_+^{0^+} \left ( \beta^{0^+} \left
( R_\uparrow + R_\downarrow \right ) - R_\downarrow \right )
\end{equation}
\begin{equation}\label{eqn9}
\mu_+^{0^+} - \mu_+^{0^-} = j \left ( \beta^{0^+} \left
(R_\uparrow - R_\downarrow \right ) + R_\downarrow \right ).
\end{equation}
These matching conditions apply for each of the four cases
illustrated in Fig.\ (\ref{fig1}).

The semiconductor near the interface is either depleted, as shown
in panels (a), (b) and (d) in Fig.\ \ref{fig1}, or accumulated as
shown in panel (c) in Fig.\ \ref{fig1}.  We input the drop in
electrostatic potential between the semiconductor side of the
interface and the edge of the depletion or accumulation region.  A
tunneling region, such as is illustrated in panel (a) in Fig.\
\ref{fig1} is described by the interface resistances $R_\uparrow$
and $R_\downarrow$, and is taken to have negligible width.  The
semiconductor side of the interface starts at the right of the
tunneling region.  From the input potential drop we calculate: the
current density; the net bias voltage, which may include a
contribution from the interface resistance; and the electrostatic
profile.  For the depleted cases we use the usual depletion
approximation to describe the electrostatics in the semiconductor.
For the accumulated case, we assume that hole current is limited
by an interfacial barrier, take a constant hole quasi-Fermi energy
in the semiconductor as shown in Fig.\ \ref{fig1}(c), and solve
Poisson's equation self-consistently to determine the
electrostatic potential in the accumulation region. Details of the
electrostatics are described in Appendix A.

In the semiconductor, electron and hole currents satisfy
continuity equations,
\begin{equation}\label{eqn10}
\frac {\partial j_\eta}{\partial x} = - e \left ( g_\eta - r_\eta
\right ),
\end{equation}
and
\begin{equation}\label{eqn11}
\frac {\partial j_p}{\partial x} =   e \left ( g_p - r_p \right ),
\end{equation}
where $g$ is a generation rate and $r$ a recombination rate.  For
spin polarized electrons there is a contribution to the
recombination rate from both spin flip scattering and
electron-hole recombination,
\begin{equation}\label{eqn12}
r_\uparrow = \frac {n_\uparrow}{\tau_r} +  \frac {n_\uparrow -
n_\downarrow}{\tau_s},
\end{equation}
where $\tau_r$ and $\tau_s$ are the recombination and spin flip
times, respectively, and an analogous expression applies for
$r_\downarrow$.  We use a drift-diffusion approximation to
describe electron and hole currents,
\begin{equation}\label{eqn13}
j_\eta = \bar \mu_n \frac {n_i}{2} kT e^{e \phi /kT} \frac
{\partial e^{\mu_\eta /kT}} {\partial x},
\end{equation}
and
\begin{equation}\label{eqn14}
j_p = -\bar \mu_p n_i kT e^{e \phi /kT} \frac {\partial e^{- \mu_p
/kT}} {\partial x}
\end{equation}
where $\bar \mu_{n (p)}$ is the electron (hole) mobility, $n_i$ is
the intrinsic carrier concentration and $\phi$ is the
electrostatic potential.  The carrier densities are given by
\begin{equation}\label{eqn15}
n_\eta = \frac {n_i}{2} e^{(e \phi + \mu_\eta)/kT}
\end{equation}
and
\begin{equation}\label{eqn16}
p = n_i e^{-(e \phi + \mu_p)/kT}.
\end{equation}

It is convenient to go into a representation describing the
electron charge and spin degrees of freedom and we define
\begin{equation}\label{eqn17}
\Omega_\pm = e^{\mu_\uparrow /kT} \pm e^{\mu_\downarrow /kT}.
\end{equation}
so that
\begin{equation}\label{eqn18}
j_\pm = \bar \mu_n \frac {n_i}{2} kT e^{e \phi /kT} \frac
{\partial \Omega_\pm} {\partial x},
\end{equation}
and
\begin{equation}\label{eqn19}
n_\pm = \frac {n_i}{2} e^{e \phi /kT}  \Omega_\pm .
\end{equation}
The corresponding generation and recombination rates are
$g_\pm$$=$$g_\uparrow$$\pm$$g_\downarrow $,
$r_+$$=$$n_+$$/$$\tau_r$, and
$r_-$$=$$n_-$$($$2$$/$$\tau_s$$+$$1$$/$$\tau_r$$)$.  The
continuity equations become
\begin{equation}\label{eqn20}
\frac {\partial j_\pm}{\partial x} = -e \left ( g_\pm - r_\pm
\right ).
\end{equation}
Substituting the drift-diffusion form into the continuity equation
gives a transport equation for $\Omega_\pm$,
\begin{equation}\label{eqn21}
\frac {\partial^2 \Omega_\pm}{\partial x^2} + \left ( \frac
{e}{kT} \frac{\partial \phi}{\partial x} \right ) \frac {\partial
\Omega_\pm}{\partial x} - \frac {1}{\Lambda_\pm^2} \Omega_\pm = -e
\tilde g_\pm e^{-e\phi/kT}
\end{equation}
where $\tilde g_\pm$$=$$g_\pm/[\bar \mu_n(kT/e)(n_i/2)]$,
$\Lambda_+^2$$=$$(kT/e){\bar \mu_n}\tau_r$, and
$\Lambda_-^2$$=$$(kT/e){\bar \mu_n}(2/\tau_s+1/\tau_r)^{-1}$. An
analogous equation holds for holes.  Analytic solutions for
$\Omega_{\pm}$ are discussed in Appendix B.

Boundary conditions at the semiconductor side of the interface,
$x$$=$$0^+$, and at the depletion edge, $x$$=$$w$, are used to
determine the two matching coefficients (see Appendix B) that
appear in the solutions of Eq.\ (\ref{eqn21}).  Details of the
boundary conditions differ somewhat for the individual cases and
will be specified in the discussion of these cases.  For the
charge degree of freedom, we use interface recombination boundary
conditions at the semiconductor side of the interface,
\begin{equation}\label{diode}
j_+ ^{0^+} =ev_{sr} \left ( n_+^{0^+} - n_{eq}^{0^+} \right )
\end{equation}
where $v_{sr}$ is the surface recombination velocity and
$n_{eq}^{0^+}$ is the equilibrium electron density at the
semiconductor side of the interface.  For the n-type semiconductor
cases we set the electron density at the depletion edge
$(x$$=$$w)$ equal to the bulk doping density so that that the
material becomes charge neutral at this point $n_+(w)$$=$$N_d$
where $N_d$ is the bulk doping density.  From the definition of
$\Omega_\pm$, we see that
\begin{equation}
\mu_- = 2kT \tanh ^{-1} \left ( \frac {\Omega_-}{\Omega_+} \right
).
\end{equation}
Combined with Eq.\ (\ref{eqn7}) this gives a boundary condition
for $\Omega_-$ at the semiconductor side of the interface.  It is
often useful to write
\begin{equation}
\frac {\Omega_-}{\Omega_+} = \frac {\frac{\partial\ln
\Omega_+}{\partial x}}{\frac{\partial\ln \Omega_-}{\partial x}}
\left ( 2\beta -1 \right )
\end{equation}
and
\begin{equation}
\left ( 2\beta -1 \right ) = \frac {j_-}{j_+} = \frac {\frac
{\partial \Omega_-}{\partial x}} {\frac {\partial
\Omega_+}{\partial x}}.
\end{equation}
This form can be useful because $(2\beta$$-$$1)$ can become the
unknown in the matching condition of Eq.\ (\ref{eqn8}).  In the
doped material beyond the depletion region the electric field is
small and spatially uniform.  In the usual treatment of current
flow in Schottky diodes this small field is neglected.  For most
of the cases there is no generation term in Eq.\ (\ref{eqn21}) and
we only need the homogenous solution for $\Omega_-$
\begin{equation}\label{diffbulk}
\Omega_-(x \geq w)=\Omega_-(w)\exp[(w-x)/\ell_+]
\end{equation}
where
\begin{equation}\label{diffbulk1}
\ell_{\pm}^{-1}=\pm \frac{e|E|}{2kT} +  \sqrt { \left( \frac
{e|E|}{2kT} \right)^2 + \left( \frac{1}{\Lambda_s}\right)^2 }.
\end{equation}
Here, $E$ is the uniform electric field in the doped material
beyond the depletion region. Substituting into Eq.\ (\ref{eqn19})
we obtain the more familiar notation of the \textquotedblleft
drift-diffusion\textquotedblright\ framework. In the bulk, $n_-$
relaxes according to
\begin{equation}\label{diffbulk2}
n_-(x \geq w)=n_-(w)\exp[(w-x)/\ell_-]
\end{equation}
where $\ell_-$ reflects the field modification of the diffusion
length at constant carrier density.\cite{pikus,yu} Matching the
continuity of $\Omega_-$ and its spatial derivative, the current
from Eq.\ (\ref{eqn1}), at the edge of the depletion region give
the final boundary conditions.

\section{calculated results of model contacts}
We discuss results for each of the four cases shown in Fig.\
\ref{fig1} sequentially.
\subsection{Injection at an n-type contact.}
\begin{figure}
\rotatebox{0}{ \includegraphics{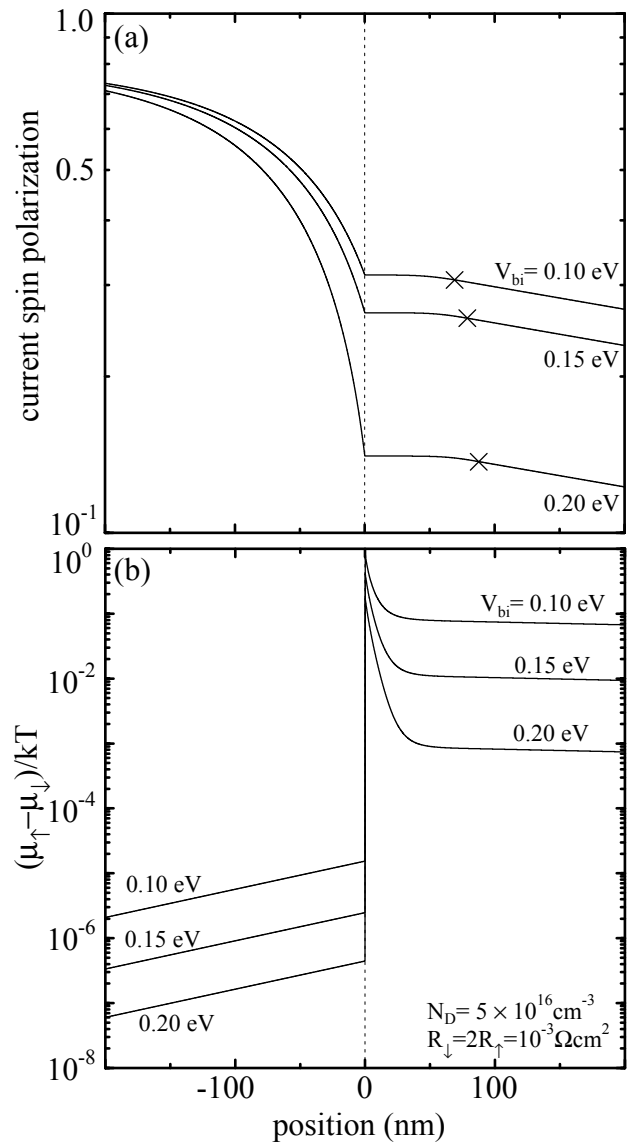}} \caption{\label{fig2}
Calculated spin injection properties for an n-type Schottky
contact as shown in Fig.\ \ref{fig1}(a). (a) Current spin
polarization $(j_-/j_+)$; and (b) electrochemical potential
difference for spin-up and spin-down electrons as a function of
position for various values of the effective Schottky barrier.
The edge of the depletion region is indicated by $\times$ on the
curves of part (a).}
\end{figure}
\begin{figure}
\rotatebox{0}{ \includegraphics{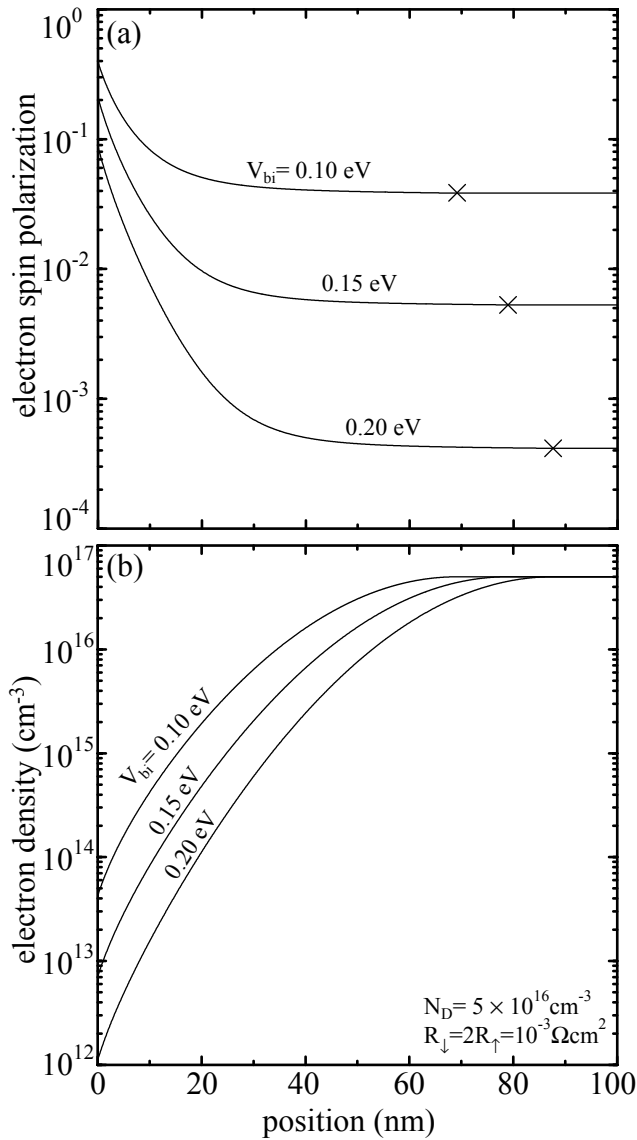}} \caption{\label{fig3}
Effect on the local spin populations of injection at an n-type
contact as shown in Fig.\ \ref{fig1}(a).  (a) Electron density
spin polarization $(n_-/n_+)$; and (b) total electron density
$(n_+)$ as a function of position for various values of the
effective Schottky barrier.  The edge of the depletion region is
indicated by the $\times$ on the curves of part (a).}
\end{figure}
\begin{figure}
\rotatebox{0}{ \includegraphics{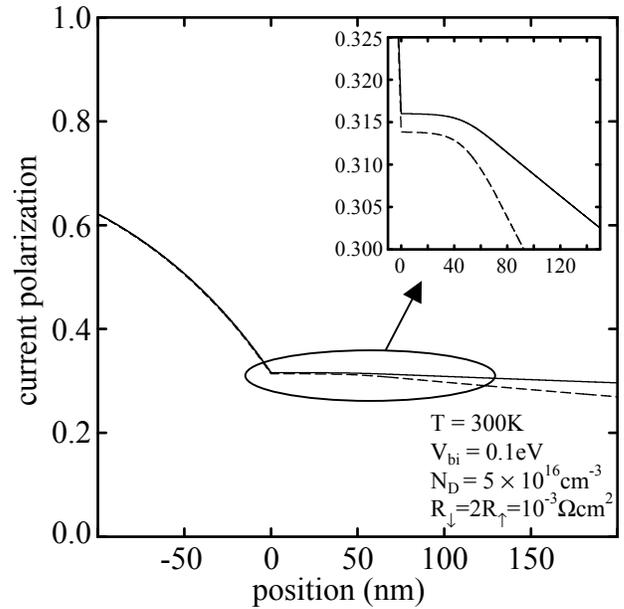}} \caption{\label{fig4}
Effect of an electric field beyond the depletion region.  The
solid curve shows current polarization as a function of position
including a greatly exaggerated electric field in the doped region
beyond the depletion region whereas the dashed curve assumes no
electric field outside the depletion region (as in Fig.\
\ref{fig2}) for $V_{bi}$$=$$0.1V$.  The field used for the solid
line was chosen to give $\ell_-$$=$$10\mu$m, or 70 times the value
determined from drift in the bulk.  If the field determined by
$en_+\bar{\mu_n}$ and $j$ is used, the result is essentially
indistinguishable from that using zero field.}
\end{figure}
This section discusses results for electron spin injection at an
n-type Schottky contact as shown in Fig.\ \ref{fig1}(a).  The
effective Schottky barrier of the of contact can be varied by
changing the doping profile in the semiconductor near the
interface as shown at the bottom of the panel.  We consider a
heavily doped region near the interface that creates a narrow
tunneling region.  The effect of the narrow doping region is to
reduce the effective Schottky barrier energy with the associated
reduction in the depletion width.  This approach to tailoring
effective Schottky barriers is well established in semiconductor
device applications.\cite{sze}

For this case, there is no hole current or optical generation.
Consequently, we solve the homogeneous form of Eq.\ (\ref{eqn21})
matched to boundary conditions at the interface given by Eqs.\
(\ref{eqn7}-\ref{eqn9}) and at the depletion edge by Eqs.\
(\ref{diffbulk}) and (\ref{eqn1}). The tunneling regions are
parameterized using interface resistances and reduced effective
Schottky barriers.  The contact is metallic with resistance equal
to $10^{-5} \Omega$ cm, polarization $\alpha_c$$=$0.9 (80\%
polarized) and spin diffusion length equal to $\Lambda_c$$=$100
nm.  The n-type semiconductor has an electron mobility of ${\bar
\mu}_n$$=$5000 cm$^2$/(V s) and a spin diffusion length equal to
1.0 $\mu$m. The diode characteristic from Eq.\ (\ref{diode}) is
determined using $v_{sr}$$=$$10^7$ cm/s.

In Figs.\ \ref{fig2}, \ref{fig3}, and \ref{fig4} we show
calculations of spin injection through a depleted n-type Schottky
contact at T=300K.  In Fig.\ \ref{fig2}, we show the effect of the
effective Schottky barrier on spin injection.  In panel (a), the
current spin polarization as a function of position is plotted for
three effective Schottky energy barriers, as labelled in the
figure.  For each barrier energy, the structure is biased to
operate at 90\% of the reverse-saturation current.  The
calculation shows clearly that the presence of an energy barrier
degrades the performance of the spin injecting structure, and that
the dependence on barrier height is strong.  In panel (b), the
corresponding electrochemical potential differences are plotted
for each structure.  We see from this panel that the origin of the
splitting in electrochemical potentials (directly related to
polarization) is from the interface resistance.  If the interface
resistance is lowered or if the $R_\uparrow/R_\downarrow$ ratio
approaches unity, then the injection properties of the structure
degrade.  Some specifics regarding the barrier lowering and
interface resistance for the n-type injector are discussed Ref.
\onlinecite{brief}.

The electron spin density polarization can be examined in the
presence of the electron density profile.  For the same conditions
used in Figs.\ \ref{fig2}, the spin polarization of the local
electron density is shown in panel (a) of Fig.\ \ref{fig3}.  The
total density is shown in panel (b) of Fig.\ \ref{fig3} for
comparison.  The polarized current may persist deeper into the
semiconductor than its ability to spin polarize the local electron
gas.  A spin polarized current may be established in the
semiconductor without strongly perturbing the spin polarization of
a background of free carriers.  However, for optical detection,
such as the spin-LED, the signal is proportional to the spin
polarization of the local density and not of the current.

In the calculations presented in Fig.\ \ref{fig2} and Fig.\
\ref{fig3} the effect of an electric field in the doped region
outside the depletion region was neglected.  This is the usual
approximation in describing the electrical properties of Schottky
diodes.  It is reasonable because the doped region outside the
depletion region is conductive and the current flow is limited by
the depletion region.  In Fig.\ \ref{fig4} we compare a
calculation of the current polarization as a function of position
neglecting the electric field in the doped region with one which
includes a greatly exaggerated value for the electric field
outside the depletion region for an effective barrier height of
0.1 eV.  The field used for the solid line was chosen to give
$\ell_-$$=$$10\mu$m and is 70 times that determined by the
conductance of this region and the injected current density. If a
field determined by the conductance and current density
($E$$=$$j/en_+\bar{\mu}$) is used, the result is essentially
indistinguishable from that using zero field. (For the calculation
in Fig.\ \ref{fig4} that field is $-$37 V/cm.) The figure shows
that for a Schottky structure with a significant effective barrier
height the electric field in the doped region outside the
depletion region has little effect on the spin injection
properties of the structure.  The reason for this is that the
matching conditions on the currents and electrochemical potentials
are at the interface between the metal and the depleted region of
the semiconductor where the concentration of electrons is
exceedingly small.  This is to be contrasted with uniform
conductivity models where the electron concentration is the same
up to the interface so that $\partial \mu_\eta/\partial x$ is
driven by the electric field on the semiconductor
side.\cite{yu,pikus}

\begin{figure*}
\rotatebox{0}{ \includegraphics{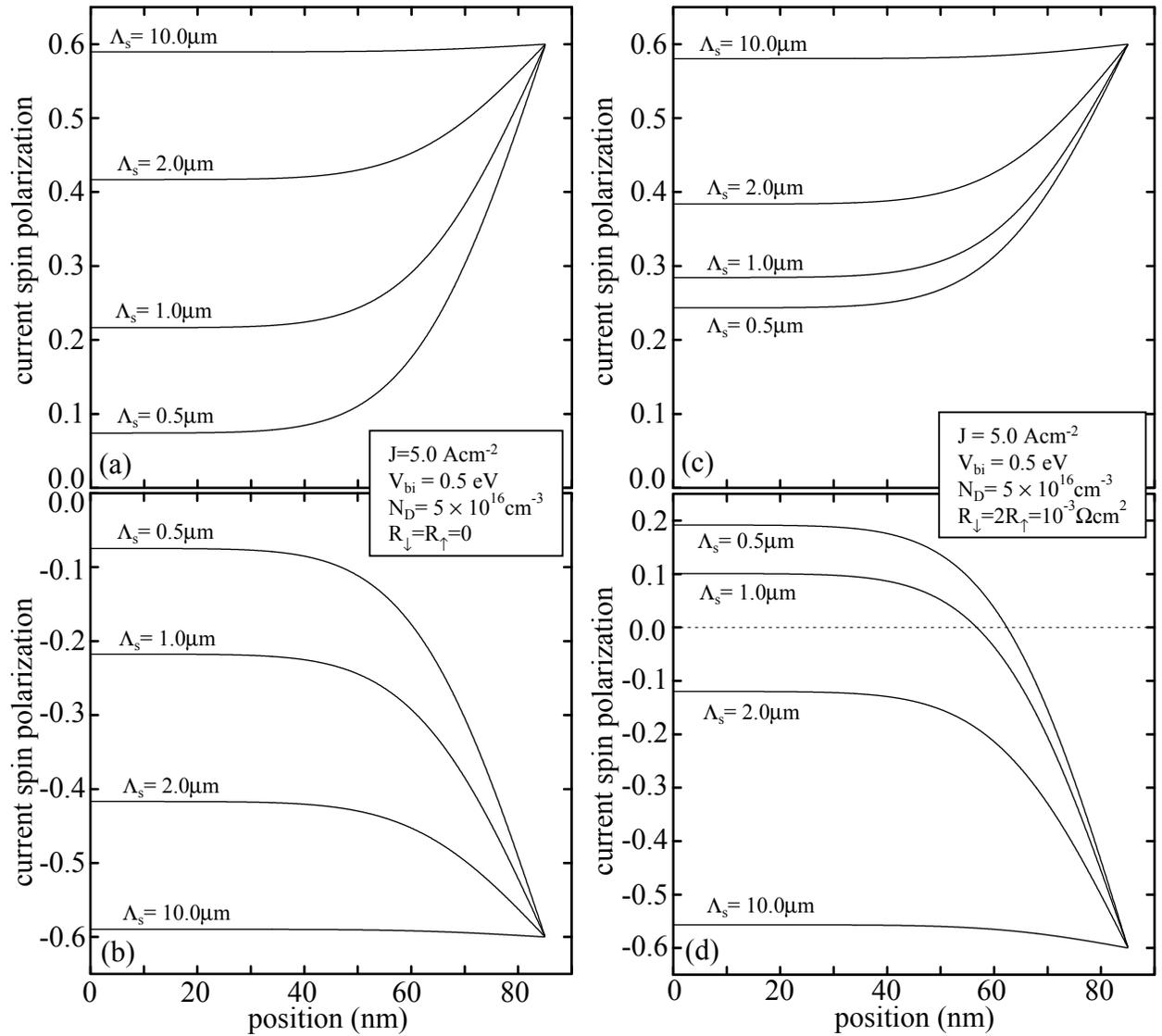}} \caption{\label{fig5}
Current spin polarization in the semiconductor for an n-type
Schottky structure operating in detection mode as shown in Fig.\
\ref{fig1}(b).  Results are plotted for various $\Lambda_s$ at
fixed V$_{\text{bi}}$.  The electron flux, incident right-to-left
at the depletion edge, corresponds to a current equal to
$\text{5.0 A cm}^{-2}$. (a) Incident polarization
$j_-(w)/j_+$$=$$0.6$ and zero interface resistance. (b)
$j_-(w)/j_+$$=$$-0.6$ and zero interface resistance. (c)
$j_-(w)/j_+$$=$$0.6$ with fixed interface resistance. (d)
$j_-(w)/j_+$$=$$-0.6$ with fixed interface resistance.}
\end{figure*}

The depleted region that occurs at at a Schottky contact is seen
to be detrimental to spin injection at a ferromagnetic metal/
semiconductor interface.  The problem arises because injection is
into a very low resistance region of the semiconductor that is
depleted of carriers.  However, the depletion region can be
tailored using a doping density profile to minimize these
deleterious effects.  For example, a heavily doped region near the
interface, such as a delta-doped layer, can be used to form a
sharp potential profile through which electrons tunnel to reduce
the effective Schottky energy barrier that determines the
magnitude of the depletion region.

\subsection{Detection at an n-type contact.}
\begin{figure*}
\rotatebox{0}{ \includegraphics{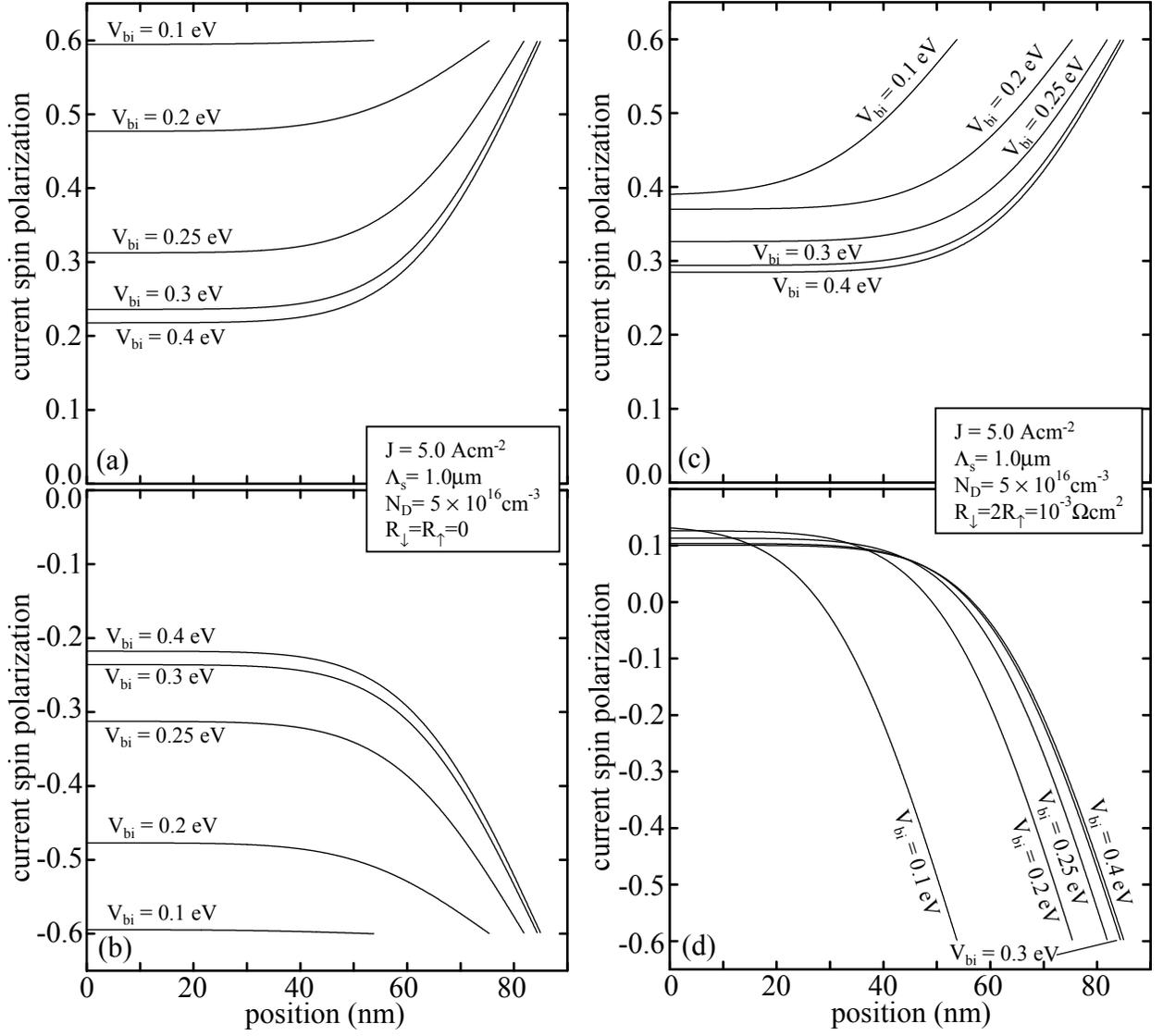}} \caption{\label{fig6}
Current spin polarization in the semiconductor for an n-type
Schottky structure operating in detection mode as shown in Fig.\
\ref{fig1}(b).  Results are plotted for various V$_{\text{bi}}$ at
fixed $\Lambda_s$.  The electron flux, incident right-to-left at
the depletion edge, corresponds to a current equal to $\text{5.0 A
cm}^{-2}$. (a) Incident polarization $j_-(w)/j_+$$=$$0.6$ and zero
interface resistance. (b) $j_-(w)/j_+$$=$$-0.6$ and zero interface
resistance. (c) $j_-(w)/j_+$$=$$0.6$ with fixed interface
resistance. (d) $j_-(w)/j_+$$=$$-0.6$ with fixed interface
resistance.}
\end{figure*}
\begin{figure}
\rotatebox{0}{ \includegraphics{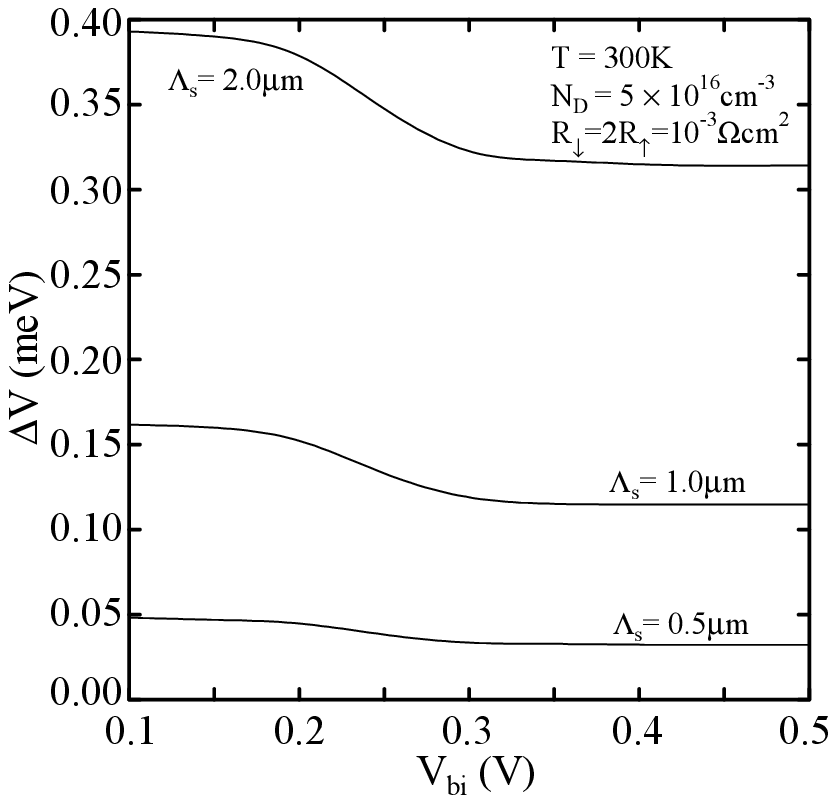}} \caption{\label{fig7}
Voltage signal as a function of effective Schottky barrier energy
for $j_-/j_+$$=$$\pm0.6$ and incident current equal to $\text{5.0
A cm}^{-2}$.  Results are shown for three values of the spin
diffusion length.}
\end{figure}
This section describes results for spin detection at n-type
Schottky structures as shown in Fig.\ \ref{fig1}(b).  The spin
detection case is similar to that of spin injection discussed in
the previous section but with a modification of the boundary
conditions.  The boundary condition on the current spin
polarization at $x$$=$$w$ is an incident polarized current
$j_-(w)$.  We consider a constant total current density of 5.0 A
cm$^{-2}$ and seek a voltage signal as the polarity of the spin
polarization is reversed.  The contact is metallic and has the
same properties as in the previous section.

The calculated current polarization as a function of position
within the depletion region for detector operation is shown in
Figs.\ \ref{fig5} and \ref{fig6}.  Results are calculated for 60\%
incident spin polarization.  In Fig.\ \ref{fig5}, we examine the
dependence of the spin polarized current on $\Lambda_s$ for fixed
effective Schottky barrier.  Since the effective Schottky barrier
is fixed the depletion width is the same for the various cases.
Long spin relaxation times result in larger spin polarizations at
the interface.  In panels (a) and (b) of Fig.\ \ref{fig5} the
current polarization behavior is almost identical for up and down
incident currents when the interface resistance is zero.  This
shows that the presence of a polarized contact material (polarized
spin up) has little impact on the spin polarized current in the
semiconductor.  In panels (c) and (d) the currents are calculated
including interface resistance and show a strong asymmetry owing
to the mismatch in $R_\uparrow$ and $R_\downarrow$.

In Fig.\ \ref{fig6} we examine the dependence of the current
polarization in the depletion region on $V_{bi}$ at fixed
$\Lambda_s$.  The bias conditions have been adjusted to give the
same total current density for all cases.  Structures with
different barrier heights have different depletion widths.  A main
point seen from Fig.\ \ref{fig6} is that large effective barrier
energies result in small spin polarizations at the interface.  In
panels (a) and (b) of Fig.\ \ref{fig6} the current polarization
behavior is very close for up and down incident currents when the
interface resistance is zero.  In panels (c) and (d) the currents
are calculated including interface resistance and show a strong
asymmetry owing to the mismatch in $R_\uparrow$ and
$R_\downarrow$.

We consider an n-type Schottky detector structure at a constant
total current density.  To fix the total current in the structure,
the forward bias ($V_F$ in Fig.\ \ref{fig1}(b)) is tuned for each
structure.  The detected signal is the change in voltage at fixed
current density when the spin polarization of the incident current
is reversed.  This voltage difference is obtained by integrating
the electrochemical potential over position from $-\infty$ to
$+\infty$ and taking the difference for two incident spin
polarizations.  After cancellations, the surviving terms yield a
voltage difference of
\begin{equation}\label{ndet}
\Delta V = \Delta \left( \frac {j_-^{0^+}}{j_+} \right) \times
\frac{j_+}{4} \left[R_\uparrow - R_\downarrow + \frac{\Lambda_c
\left( 1-2\alpha_c \right)}{\sigma_c \alpha_c \left( 1 - \alpha_c
\right)} \right]
\end{equation}
where $\Delta$ indicates the difference between the quantities for
opposite signs of the incident spin polarization at $x$$=$$w$.

>From Eq.\ (\ref{ndet}) we see that without a spin dependent
interface resistance ($R_\uparrow$$=$$R_\downarrow$$\rightarrow$0)
and for a highly conductive contact
($\sigma_c$$\rightarrow$$\infty$) no significant voltage
difference can be established.  Using metallic contacts
($\sigma_c$$=$$10^{5}$ $\mho$ cm$^{-1}$) with contact spin
diffusion lengths less than a micron ($\Lambda_c$$=$10$^{-5}$ cm)
and currents ($\sim$1 A cm$^{-2}$) corresponding to low biasing
conditions the calculated detected voltage differences are
negligibly small ($\lesssim$10$^{-10}$ V).  We conclude that
Schottky contacts without a spin-selective tunnel barrier will not
be useful as spin polarized current detectors and therefore we
concentrate on Schottky structures containing a spin selective
interface resistance.  In Fig.\ \ref{fig7} we show calculated
voltage differences as a function of effective Schottky barrier
for three values of the spin-diffusion length in the
semiconductor.  The detection signal saturates for both large and
small $V_{bi}$ values.  For small barriers the depletion region
vanishes and the incident polarized current reaches the interface
and only small applied bias is required to establish the constant
current.  For large depletion widths, larger applied biases are
required to keep the current density fixed and the resulting
resulting interface current polarization saturates consistent with
the behavior shown in Fig.\ \ref{fig6}(c) and \ref{fig6}(d).

The depleted region that occurs at at a Schottky contact is seen
to be detrimental to spin detection at a ferromagnetic metal/
semiconductor interface.  The problem arises because in these
forward biased structures, electron current is driven by diffusion
against a strong and rapidly varying electric field in the
depletion region.  As a result the effective drift-diffusion
lengths in the depletion region can become rather short leading to
strong spin relaxation.  As for the electron injection structures
the depletion region can be tailored using a doping density
profile to minimize these deleterious effects.

\subsection{Injection at a p-type contact.}

\begin{figure}[b]
\rotatebox{0}{ \includegraphics{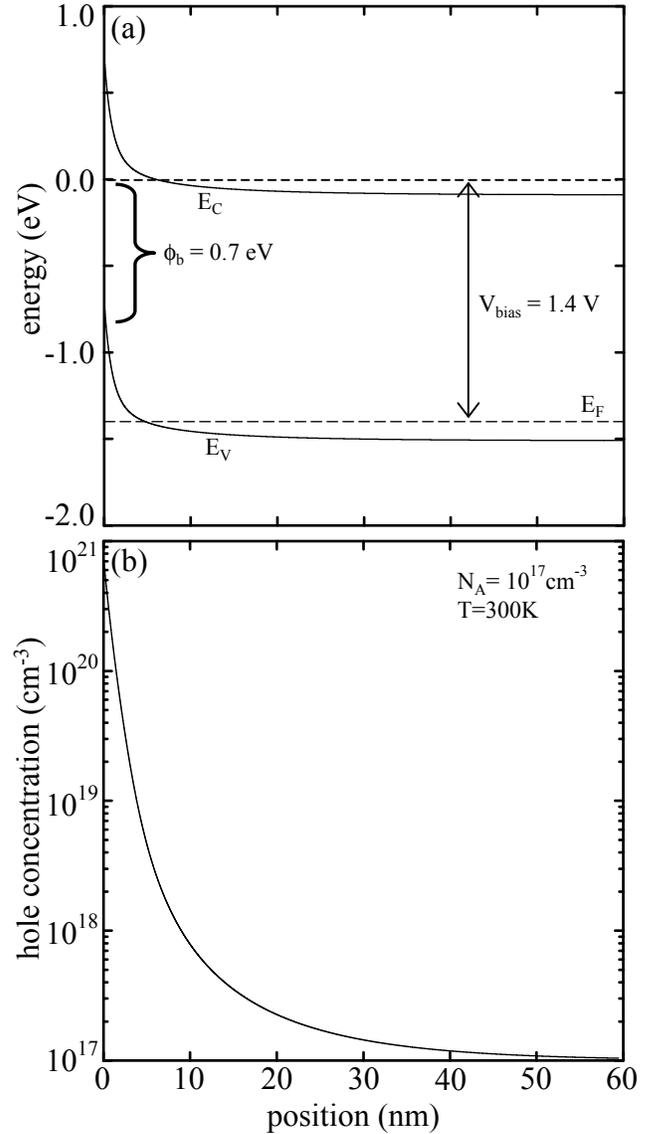}} \caption{\label{fig8}
Energy band diagram (a) and hole density profile (b) of an
accumulated p-type Schottky diode in strong forward bias as in
Fig.\ \ref{fig1}(c).}
\end{figure}

\begin{figure}
\rotatebox{0}{ \includegraphics{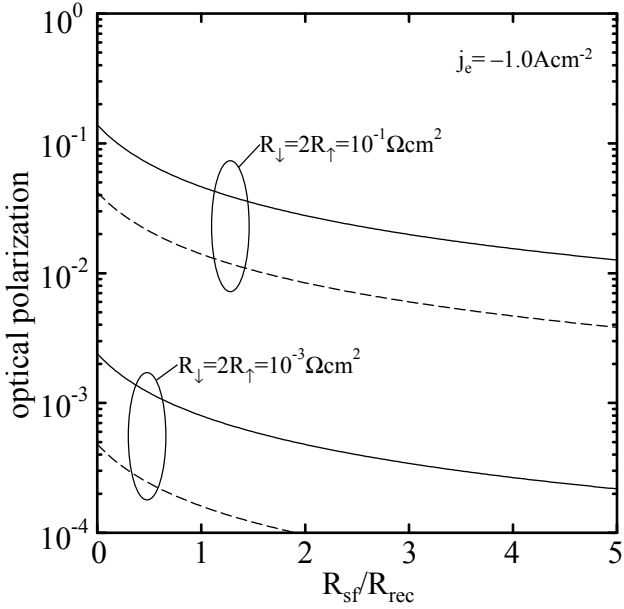}} \caption{\label{fig9}
Integrated emitted photon polarization
$(\sigma_+$$-$$\sigma_-)$$/$$(\sigma_+$$+$$\sigma_-)$ as a
function of electron spin-flip scattering rate coefficient for
fixed injected current density from a ferromagnetic contact
$(\alpha_c$$=$$0.9)$ into the accumulated p-type contact shown in
Fig.\ \ref{fig8}.  Results are shown for injection efficiencies of
$j_+/j$$=$0.5 (solid curves) and 0.1 (dashed curves).}
\end{figure}

This section describes results for spin polarized electron
injection from a ferromagnetic contact into a forward biased
p-type Schottky diode.  The hole current is limited by a barrier
at the interface but is most likely larger that the minority
electron injection current.  The output signal is the ratio of
right to left circularly polarized light emitted from the
semiconductor when the injected electrons recombine radiatively
with unpolarized holes in the p-type material.

The electrostatic treatment for the p-type structure is described
in Appendix A.  Calculated energy band and hole concentration
profiles are shown in Fig.\ \ref{fig8} and correspond to the
semiconductor portion of Fig. \ref{fig1}(c).  An interfacial
barrier is an important feature of the structure to prevent
runaway hole current at the interface.  As a consequence of the
barrier there is strong accumulation of holes near the interface.
The large and rapidly varying hole concentration requires a
different treatment of the electron spin relaxation
process\cite{bap} than used in the depletion cases considered
above. To account for the presence of the accumulated holes, we
use the local hole concentration to vary the recombination and
spin-flip scattering rates as a function of position.

We take the electron spin relaxation times to be linear with the
local hole density and define rate coefficients $R_{sf}$ and
$R_{rec}$ for spin-flip scattering and recombination so that
$\tau_s^{-1}$$=$$R_{sf}$$\times$$p(x)$ and
$\tau_r^{-1}$$=$$R_{rec}$$\times$$p(x)$.  We present calculations
for a range of $R_{sf}$.  We take
$R_{rec}$$=$7$\times$$10^{-10}$cm$^3$ s$^{-1}$, a typical value
for p-type GaAs at T=300K.\cite{rate} The hole mobility is ${\bar
\mu}_p$$=$500 cm$^2$/(V s).

Unlike for the depleted structures there is no closed form
analytic solution for the spin-dependent transport equations for
the accumulated structure.  Given a numerical solution for $p(x)$
from the electrostatic calculation, the solution of Eq.\
(\ref{eqn21}) is obtained by numerical integration and the
shooting criterion that solutions be non-diverging as
$x$$\rightarrow$$\infty$.  There is no optical generation of
carriers ($g_\pm$$=$$0$) so that only the homogeneous solution is
required.

The detected quantity in this case is the degree of circular
polarization of the emitted light.  Assuming good radiative
recombination efficiency, the optical polarization of right and
left circularly polarized light ($\sigma_\pm$) is proportional to
the local electron density and is given by
\begin{equation}
\frac{\sigma_+ - \sigma_-}{\sigma_+ + \sigma_-}=\frac
{\int_{0}^{\infty}n_{-}dx} {\int_{0}^{\infty}n_{+}dx} = \frac
{\Lambda_{-}^{2}} {\Lambda_{+}^{2}} \left(2\beta^{0^+} -1\right).
\end{equation}
The spin polarization ($2\beta^{0^+}$$-$$1$) of the electron
current at the interface is obtained from the numerical solution
of Eq.\ \ref{eqn21}.  The integration is straightforward because
the electron concentrations are related to the current through the
continuity equations, Eqs. (\ref{eqn10}-\ref{eqn12}).

The calculated optical polarizations for injection into p-type
material are shown in Fig.\ \ref{fig9} as a function of the
spin-flip scattering rate coefficient.  The curves are calculated
for the same electron current density at $x$$=$$0$.  We consider
insulating barriers that result in minority carrier injection
efficiencies equal to 50\% and 10\%.  If used as a
characterization tool, the structure should to be sensitive to
differences in the optical polarization signals in order to
determine the interface resistance values.  This is the case for
the case of slow spin-flip scattering compared radiative
recombination.  If the spin-flip scattering rate is faster than
the recombination rate in the semiconductor  characterization by
this method is will to be difficult.

\subsection{Detection at a p-type contact.}
\begin{figure}[b]
\rotatebox{0}{ \includegraphics{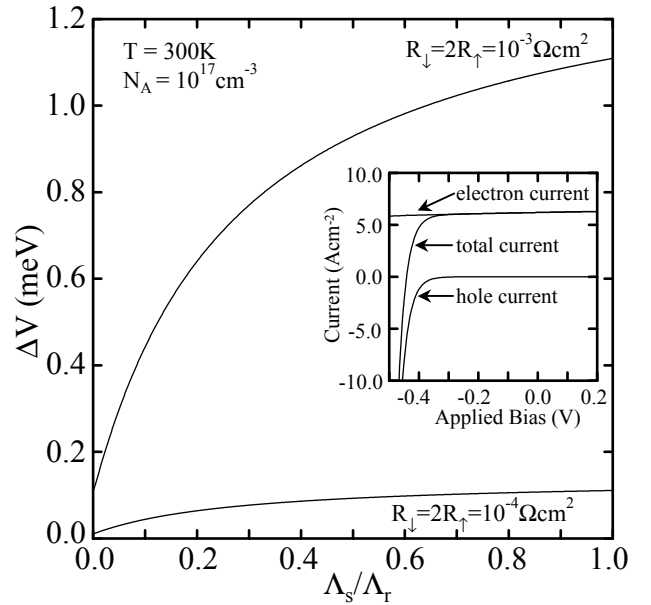}} \caption{\label{fig10}
Voltage difference for collection of spin polarized electrons at a
depleted p-type contact.  The inset shows the electron and hole
current contributions to the current-voltage characteristics.}
\end{figure}

The p-type detector structure shown in Fig.\ \ref{fig1}(d)
involves optical generation of carriers.  The p-type Schottky
diode is at zero or small bias (either forward or reverse).  As in
the depleted n-type detector, we compute the voltage difference
established as the circular polarization of the light is changed
from  right to left.  We assume that the optical generation is
nearly uniform over the depletion depth (i.e., that the reciprocal
of the absorption coefficient is small compared to the depletion
width) and that the optical generation of spin-polarized carriers
is a 3:1 or 1:3 ratio of spin-up to spin-down electrons according
to the optical selection rules. Beyond the depletion edge, for
$x$$>$$w$, the optical generation falls off as $\exp(-\chi x)$
with $\chi$$=$$10^4$ cm. Under these conditions, the transport
equation (\ref{eqn21}) is solved analytically as described in
Appendix B.

A calculated current/voltage characteristic for the various
current components calculated for this structure are shown in the
inset of Fig.\ \ref{fig10}.  The properties of the p-type Schottky
are those used in the accumulation case of the previous section.
The barrier height is 0.7eV and the incident light power is set to
1 W cm$^{-2}$ or $g_+$$\approx$$4.4$$\times$$10^{22}$ s$^{-1}$
cm$^{-3}$ for photons at the bandgap of GaAs.  In this structure,
the electron current stays nearly constant over a broad range of
applied bias around $V_A$$=$$0$ so we report results for zero
bias. The value for $\Delta V$, as for the n-type detector
structure, is given by Eq.\ (\ref{ndet}) except that the total
electron current at the interface $j_+$ must be accounted for
separately from the total current which includes a contribution
from holes.  This separation is straightforward once solutions for
$\Omega_{\pm,p}$ have been calculated.  In Fig.\ \ref{fig10} we
plot the calculated voltage differences as a fucntion of
$\Lambda_s$ for optical excitation.  If the semiconductor spin
lifetime has been determined by other experimental means, this
represents a second method of characterizing the interface
resistance.  Notice that even when the spin lifetime becomes
maximally long ($\Lambda_s$$=$$\Lambda_r$) there is a strong
dependence of the measured voltage on the interfacial conditions.

\section{Summary}

We have presented a theoretical description of the injection and
detection of spin polarized electrons at n-type and p-type
Schottky contacts.  The presence of the depletion region that
occurs at Schottky contacts has been shown to be detrimental to
both spin injection and detection in n-type Schottky structures.
The problem for the reverse biased n-type spin injection
structures arises because the injection is into a very low
resistance region of the semiconductor that is depleted of
carriers.  The problem for the forward biased n-type spin
detection structures arises because electron current is driven by
diffusion against a strong and rapidly varying electric field in
the depletion region.  As a result the effective drift-diffusion
lengths in the depletion region can become rather short leading to
strong spin relaxation in the depletion region.

For both n-type injection and detection structures, the depletion
region can be tailored using a doping density profile to minimize
these deleterious effects.  A heavily doped region near the
interface, such as a delta-doped layer, can be used to form a
sharp potential profile and this tunneling region effectively
reduces the Schottky energy barrier that determines the magnitude
of the depletion region.  The model results indicate that
efficient spin-injection and spin-polarization detection can be
achieved in these n-type structures if they are properly designed
so that the effective Schottky barrier is reduced to less than
about 0.2 eV.

We discussed two experimental cases in which ferromagnetic
Schottky contacts to p-type semiconductors could be used to
characterize the spin dependent transport properties of interface
tunnel barriers: optical detection of spin currents injected into
a strongly forward biased accumulated p-type semiconductor and
electrical measurement of optically excited spin populations at
zero biased p-type contacts.  The same structure can be used under
different bias and excitation conditions for the two experiments.

A set of spin-labelled transport equations has been developed, in
a systematic way, that is suitable for device models at a basic
level.  We have demonstrated that the electrostatic and current
conditions that are present in actual devices can lead to
important consequences for spin dependent transport in the
structures must be taken into account.  With these building blocks
in place, there are clear extensions of the model to more complex
device structures which will be the focus of future work.

\begin{acknowledgments}
This work was supported by the SPINs program of the Defense
Advance Research Projects Agency.
\end{acknowledgments}

\appendix
\section{schottky contact electrostatics}

The band-bending that occurs near a Schottky contact is central to
the discussion of spin transport in these structures.  This
appendix describes the electrostatic inputs to the solutions of
the spin transport equations.

Three of the cases, shown in Fig.\ \ref{fig1}(a), (b), and (d),
involve a contact under small bias.  In these cases, we use the
depletion approximation where the electrostatic potential and
depletion width are given by
\begin{equation}
\phi(x)=\phi(0)\pm \frac{eN}{\varepsilon_s \varepsilon_o}wx \mp
\frac{1}{2}\frac{eN}{\varepsilon_s \varepsilon_o}x^2
\hspace{0.5cm} 0\leq x\leq w
\end{equation}
and
\begin{equation}
w=\sqrt{\frac{2\varepsilon_s
\varepsilon_o}{eN}\left(V_{bi}+V_A\right)}
\end{equation}
here $\varepsilon_o$ is the free space permittivity, and
$\varepsilon_s$ is the relative static dielectric constant of the
semiconductor.  The applied voltage $V_A$ can be either sign but
must be small enough in forward bias so as not to invert the
semiconductor from depletion to accumulation.  For an n-type
contact, $N$ is the donor concentration and the upper sign
applies.  The acceptor concentration and the lower sign are used
for p-type contacts.

The inverted p-type contact shown in Fig.\ \ref{fig1}(c), requires
a numerical treatment of the electrostatics.  We consider a case
of strong forward bias in which a barrier layer limits hole
transport at the metal/semiconductor interface and we treat the
accumulation region in the semiconductor as a quasi-equilibrium
system characterized by a Fermi energy for holes with a valence
band density of states $N_V$.  The electrostatics in the
accumulation region is determined by the hole density which is
given by statistics:
\begin{equation}
p\left( x \right) =N_V\frac{2}{\sqrt{\pi}}{\cal F}_{\frac{1}{2}}
\left[\lambda_F(x) \right]
\end{equation}
where the Fermi one-half integral is given by
\begin{equation}
{\cal F}_{\frac{1}{2}}\left[y\right]=\int_{0}^{\infty
}\frac{\lambda ^{\frac{1}{2}}}{1+e^{\lambda -y}}d\lambda
\end{equation}
and
\begin{equation}
\lambda \left( x\right) =\frac{\varepsilon -E_{V}\left( x\right)
}{kT}\hspace{0.5cm}\lambda _{F}\left( x\right)
=\frac{E_{F}-E_{V}\left( x\right) }{kT}.
\end{equation}
Far from the interface the hole density is equal to the acceptor
doping level $p(\infty)$$=$$N_A$ so the local hole concentration
can be written as
\begin{equation}
p\left( x \right) =N_A\frac{{\cal F}_{\frac{1}{2}}
\left[\lambda_F(x) \right]}{{\cal F}_{\frac{1}{2}}
\left[\lambda_F(\infty) \right]}.
\end{equation}

To solve Poisson's equation, we compute the charge density by
subtracting the background density of ionized accepters which
yields
\begin{eqnarray}\label{density}
\rho(x)&=&e\left[ p\left( x\right) -N_{A}\right] \nonumber \\
&=&eN_A\frac{{\cal F}_{\frac{1}{2}} \left[\lambda_F(x)
\right]-{\cal F}_{\frac{1}{2}} \left[\lambda_F(\infty)
\right]}{{\cal F}_{\frac{1}{2}} \left[\lambda_F(\infty) \right]}.
\end{eqnarray}
These algebraic steps are made so that the numerator can be
rearranged to give an integral with a closed-form solution.  First
we rearrange the difference of Fermi integrals in the numerator
as:
\begin{multline}
{\cal F}_{\frac{1}{2}} \left[\lambda_F(x) \right]-{\cal
F}_{\frac{1}{2}} \left[\lambda_F(\infty) \right] \\
=\int_{0}^{\infty }\frac{\lambda ^{1/2}}{1+e^{\lambda
-\lambda_F(\infty)}}{\cal A} d\lambda
\end{multline}
where
\begin{eqnarray}
{\cal A}&=&\frac{1-e^{\lambda_F(\infty)-\lambda _{F}\left(
x\right)}}{e^{\lambda_F(\infty)-\lambda}
+e^{\lambda_F(\infty)-\lambda _{F}\left( x\right)} } \nonumber \\
&=&\frac{1-e^{\frac{E_V(x)-E_V\left(\infty\right)}{kT}}}{e^{\lambda_F(\infty)-\lambda}
+e^{\frac{E_V(x)-E_V\left(\infty\right)}{kT}} }.
\end{eqnarray}
We substitute the charge density into the integral form of
Poisson's equation for the electric field $F$.  This process
begins with Poisson's equation in the form
\begin{equation}\label{poisson}
\int_{F(x)}^{F\left( \infty \right)
=0}FdF=\int_{E_V(x)}^{E_V\left( \infty \right)}\frac{\rho \left(
E_V \right) }{\varepsilon_s \varepsilon_o}dE_V
\end{equation}
and results in
\begin{multline}\label{stat1}
F^{2}\left( x\right) =  \frac{2N_A}{\varepsilon_s \varepsilon_o
{\cal F}_{\frac{1}{2}} \left[\lambda_F(\infty) \right]} \\
         \times \int_{0}^{\infty }\frac{\lambda
^{1/2}}{1+\exp \left( \lambda -\lambda _{\infty }\right) }\left(
\int_{E_V(x)}^{E_V\left( \infty \right) }{\cal A}dE_V \right)
d\lambda
\end{multline}
which can be simplified by substituting the analytic expression
for the interior integral.  After substitution and simplification,
we arrive at the following result:
\begin{multline}\label{stat}
    F^{2}\left( x\right) =  \frac{2kTN_A}{\varepsilon_s \varepsilon_o
{\cal F}_{\frac{1}{2}} \left[\lambda_F(\infty) \right]}\\
    \times \int_{0}^{\infty }\frac{
\lambda ^{1/2}}{1+q}\left( {\bar q}q+\left( 1+q\right) \ln \left[
\frac{1+qe^{-{\bar q}} }{1+q}\right] \right) d\lambda
\end{multline}
where $\bar{q}$ and $q$ are defined by
\begin{equation}
{\bar q}=\frac{E_V\left( \infty \right) -E_V\left( x\right) }{kT}
\hspace{0.5cm} q=\exp \left( \lambda -\lambda _{\infty }\right) .
\end{equation}

An electrostatic profile is computed by first obtaining the
electric field at $x$$=$$0$.  Given the energy barrier (fixes
$\bar{q}$ at $x$$=$$0$) and doping density (fixes the fermi level
relative to the band edge in the bulk), we can obtain $F(0)$ with
one numerical integration of Eq.\ (\ref{stat}).  The valence band,
electric field, and density profiles are then generated by
integrating Eqs.\ (\ref{poisson}) and (\ref{stat}) forward in $x$
and evaluating Eq.\ (\ref{density}) at each spatial step.  A
sample result of this process is plotted in Fig.\ \ref{fig8}.

\section{SOLUTIONS TO THE TRANSPORT EQUATION IN THE DEPLETION REGION}

In this appendix we present analytic solutions for the transport
equation (\ref{eqn21}) when the depletion approximation is used
for the electrostatics of the structure.

In the depletion mode cases, we have a second order differential
equation in the depletion region that is solved subject to
boundary conditions at the metal/semiconductor interface and at
the depletion region edge in the semiconductor.  A typical
equation, including the possibility of optical generation, can be
written as
\begin{equation}
\frac{d^{2}\Omega}{dx^{2}}+\frac{e}{kT}\frac{d\phi }{dx}
\frac{d\Omega}{dx}-\frac{1}{\Lambda
^{2}}\Omega=-Ge^{\frac{-e\phi}{kT}}.
\end{equation}
Within the depletion approximation, we have a quadratic form of
the electrostatic potential
\begin{equation}
\frac{e\phi}{kT}=\frac{1}{2}ax^2+bx+\frac{e\phi_o}{kT}
\end{equation}
The particular solution $\Omega_{p}$ is
\begin{equation}
\Omega_{p}=\frac{G}{a+\frac{1}{\Lambda^2}}e^{\frac{-e\phi}{kT}}
\end{equation}
and the homogeneous differential equation can be written as
\begin{equation}\label{dq}
\frac{d^{2}\Omega}{dx^{2}}+\left(ax+b\right)
\frac{d\Omega}{dx}-\frac{1}{\Lambda ^{2}}\Omega=0.
\end{equation}

Solutions results from a change of variables that transform the
homogeneous differential equation to the confluent hypergeometric
equation,
\begin{equation}\label{che}
z\frac{d^{2}f}{dz^{2}}+\left(\xi-z\right)\frac{df}{dz}-\zeta f=0,
\end{equation}
which has independent solutions $M\left(\zeta,\xi,z\right)$ and
$U\left(\zeta,\xi,z\right)$, the confluent hypergeometric
functions of the first and second kinds.\cite{stegun}  In order to
avoid the use of complex coefficients (this may arise if a single
form of the solution is used with both directions of the electric
field), we give the variable substitutions and homogeneous
solutions $\Omega_h$ for n-type ($a$$<$$0$, $b$$>$$0$) and p-type
($a$$>$$0$, $b$$<$$0$) cases separately.

The necessary transformations are obtained by substituting
\begin{equation}\label{tran1}
\Omega_h=\left(ax+b\right)f(z) \hspace{0.5cm}
z=\frac{-\left(ax+b\right)^2}{2a} \hspace{0.4cm} \text{(n-type)}
\end{equation}
or
\begin{equation}\label{tran2}
\Omega_h=\left(ax+b\right)e^{-z}f(z) \hspace{0.3cm}
z=\frac{\left(ax+b\right)^2}{2a} \hspace{0.4cm} \text{(p-type)}
\end{equation}
into Eq.\ (\ref{dq}).  The result of the transformation is Eq.\
(\ref{che}) with $\xi$$=$$\frac{3}{2}$ and
\begin{equation}
\zeta=\frac{1}{2}-\frac{1}{2a\Lambda ^{2}} \hspace{0.5cm}
\text{(n-type)}
\end{equation}
or
\begin{equation}
\zeta=1+\frac{1}{2a\Lambda ^{2}} \hspace{0.5cm} \text{(p-type)}.
\end{equation}
In both the n-type and p-type solutions,
\begin{equation}
f\left(z\right)=\gamma_1M\left(\zeta,\xi,z\right)+
\gamma_2U\left(\zeta,\xi,z\right)
\end{equation}
where $\gamma_1$ and $\gamma_2$ are coefficients determined by
applying boundary conditions.

Boundary conditions applied at the depletion edge where
$x$$=$$w$$=$$-$$b/a$ require the evaluation of the special
functions as the argument vanishes, $z$$\rightarrow$$0$.  The
$M\left(\zeta,\xi,z\right)$ function approaches unity as $z$
vanishes.  The $U\left(\zeta,\xi,z\right)$ function is more
complicated.  The small argument behavior of interest is given by
\begin{multline}
\lim_{z\rightarrow 0}U(\zeta,\xi,z) \sim \frac{\pi }{\sin
\left(\pi \xi\right)} \\
\times \left\{ \frac{1}{\Gamma \left( 1+\zeta-\xi\right) \Gamma
\left( \xi\right) }-\frac{1}{z^{\xi-1}\Gamma \left( 2-\xi\right)
\Gamma \left( \zeta\right) }\right\} .
\end{multline}
The apparent singularity is cancelled by the leading factors of
$(ax$$+$$b)$ in Eqs.\ (\ref{tran1}) and (\ref{tran2}).  It can be
shown after some algebra that the functions and derivatives have
well behaved values at the depletion edge given by
\begin{equation}
\Omega_h\left( w\right) =\gamma _{2}\frac{\sqrt{-2\pi a}}{\Gamma
\left( \frac{1}{2}-\frac{1}{2a\Lambda^2}\right) } \hspace{0.3cm}
\text{(n-type)}
\end{equation}
or
\begin{equation}
\Omega_h\left( w\right) =-\gamma _{2}\frac{\sqrt{2\pi a}}{\Gamma
\left(1+\frac{1}{2a\Lambda^2}\right) } \hspace{0.3cm}
\text{(p-type)}
\end{equation}
and
\begin{equation}
\left.\frac{d\Omega_h}{dx}\right|_{x=w}=a\gamma _{1}+\gamma _{2}\left[ \frac{-2a\sqrt{%
\pi }}{\Gamma \left( \frac{1}{-2a\Lambda^2}\right) }\right]
\hspace{0.3cm} \text{(n-type)}
\end{equation}
or
\begin{equation}
\left.\frac{d\Omega_h}{dx}\right|_{x=w}=a\gamma _{1}-\gamma _{2}\left[ \frac{2a\sqrt{%
\pi }}{\Gamma \left(\frac{1}{2} +\frac{1}{2a\Lambda^2}\right)
}\right] \hspace{0.3cm} \text{(p-type)}.
\end{equation}

The other matching solutions occur for the interface at
$x$$=$$0^+$ and require only the evaluation of the special
functions for typical arguments with no special considerations.

\end{document}